\documentclass[11pt,table]{article}
\pdfoutput=1

\usepackage{authblk}
\usepackage[margin=1in]{geometry}
\usepackage[english]{babel}
\usepackage[utf8x]{inputenc}

\usepackage{color}
\usepackage{amsmath, amsfonts, bbm}
\usepackage{hyperref}
\usepackage[textsize=scriptsize,textwidth=2.2cm]{todonotes}
\usepackage{subfig}

\usepackage{natbib}
\usepackage[title]{appendix}

\title{Do Offline Metrics Predict Online Performance in \\
Recommender Systems?}

\author[1]{Karl Krauth\thanks{karlk@berkeley.edu}}
\author[1,2]{Sarah Dean}
\author[1,2]{Alex Zhao}
\author[1,2]{Wenshuo Guo}
\author[1,2]{Mihaela Curmei}
\author[1]{Benjamin Recht}
\author[1]{Michael I. Jordan}
\affil[1]{University of California, Berkeley}
\affil[2]{Equal contribution}

\date{\today}

\usepackage{xspace}

\newcommand{\reclab}{\textsf{RecLab}\xspace} 
\DeclareUnicodeCharacter{2212}{-}

\newcommand{\clip}{\mathsf{clip}}
\usepackage{subfig}

\begin{document}

\maketitle

\begin{abstract}

Recommender systems operate in an inherently dynamical setting. Past recommendations influence future behavior, including which data points are observed and how user preferences change. However, experimenting in production systems with real user dynamics is often infeasible, and existing simulation-based approaches have limited scale. As a result, many state-of-the-art algorithms are designed to solve supervised learning problems, and progress is judged only by offline metrics. In this work we investigate the extent to which offline metrics predict online performance by evaluating eleven recommenders across six controlled simulated environments. We observe that offline metrics are correlated with online performance over a range of environments. However, improvements in offline metrics lead to diminishing returns in online performance.
Furthermore, we observe that the ranking of recommenders varies depending on the amount of initial offline data available. We study the impact of adding exploration strategies, and observe that their effectiveness, when compared to greedy recommendation, is highly dependent on the recommendation algorithm. We provide the environments and recommenders described in this paper as \reclab: an extensible ready-to-use simulation framework at this URL: \url{https://github.com/berkeley-reclab/RecLab}.

\end{abstract}

\section{Introduction}
\label{sec:intro}
Recommender systems operate in dynamical settings.
The recommendations given during one round of user interaction will affect the observations used to make recommendations in the future.
The sequential nature of the problem is  complicated by the fact that a deployed recommender system contends with changing preferences, due to external causes or induced by the recommendations themselves.
From well-known effects like popularity bias in item recommendations to contested phenomena like polarization and radicalization among users; myopic optimization of offline metrics can cause unintended consequences \cite{dandekar2013polarization, abdollahpouri, faddoul2020longitudinal}.
Foreseeing the effects of deployed recommender systems is a complex socio-technical problem, depending on human psychology and behavioral economics.
But even the basic questions of reliability and reproducibility for recommendation algorithm design remain unanswered.

In this paper, we focus on the evaluation of recommender performance in dynamical settings.
Despite the fact that the recommender systems community is well aware of the challenges posed by dynamical interactions \cite{kouki2019offline}, most recommendation algorithms and preference models are primarily designed and evaluated in an offline setting \cite{lee2016llorma, sedhain2015autorec, zheng2016neural, wang2006unifying, rendle:tist2012, steck2019embarrassingly, ning2011slim}. The typical offline-first evaluation methodology involves the following four steps: 
\begin{enumerate}
 \item \textbf{Dataset Creation - } An organization or research group creates a dataset by collecting user interactions with a set of items hosted on an internet platform. Two prominent examples are the Netflix Prize and MovieLens datasets.
 \item \textbf{Offline Evaluation - } Algorithm developers use the datasets from step (1) to evaluate their recommender systems. The developers train their algorithm on a train split of the datasets, tune the algorithm's parameters on a validation split, and evaluate the algorithm on a test split using various offline metrics. Common metrics include root mean squared error (RMSE) and normalized discounted cumulative gain (nDCG).
 \item \textbf{Comparison with Baselines - } The developers compare their results on offline metrics with other algorithm's results, either by running the other algorithms themselves or by referring to previously published work. If their algorithm compares favorably the developers may publish their work, or if they can, proceed to the next step.
 \item \textbf{Online Evaluation - } The algorithm is deployed on a real platform and its performance is evaluated using online metrics, which are usually defined to capture some notion of utility. Examples of popular online metrics include click-through rate (CTR) and watch time.
\end{enumerate} 
Offline evaluations make sense given the difficulty of evaluating algorithms online. Most researchers do not have access to a platform on which to perform online evaluations, and even those that do may not be able to perform large-scale evaluations due to the engineering effort required or the potential for lost revenue.
However, offline evaluation comes with its own set of challenges. The data collected in step (1) is influenced by the recommender that is deployed at the time, which often leads to selection bias \cite{schnabel2016recommendations}.
\citet{dacrema2019we} demonstrate that step (2) of this evaluation procedure is often performed incorrectly, leading to non-reproducible results due to practices like inconsistent dataset splitting. 
Furthermore, they show that the baselines in step (3) are often incorrectly tuned, leading to a false sense of progress, a finding corroborated by \citet{rendle2019difficulty}.
The difficulty of properly evaluating algorithms offline brings into question the relationship between steps (1) to (3) and step (4).


In this work, we study the relationship between offline and online evaluation in controlled simulation environments to see the impact of recommender feedback effects, user dynamics, exploration, and low data. 
By using large-scale simulations, our results isolate these effects devoid of any confounding factors.
We evaluate eleven recommendation algorithms across six simulated environments.

The recommenders we evaluate encompass simple baselines, neighborhood-based models, kernel-based models, linear models, factorization models, and neural models. The simulated environments on which we evaluate represent a diverse set of scenarios including two settings implemented in prior work and one setting that is initialized using the MovieLens dataset \cite{harper2015movielens}.

\begin{figure*}[ht]
    \centering
    \subfloat{\includegraphics[width=0.45\textwidth]{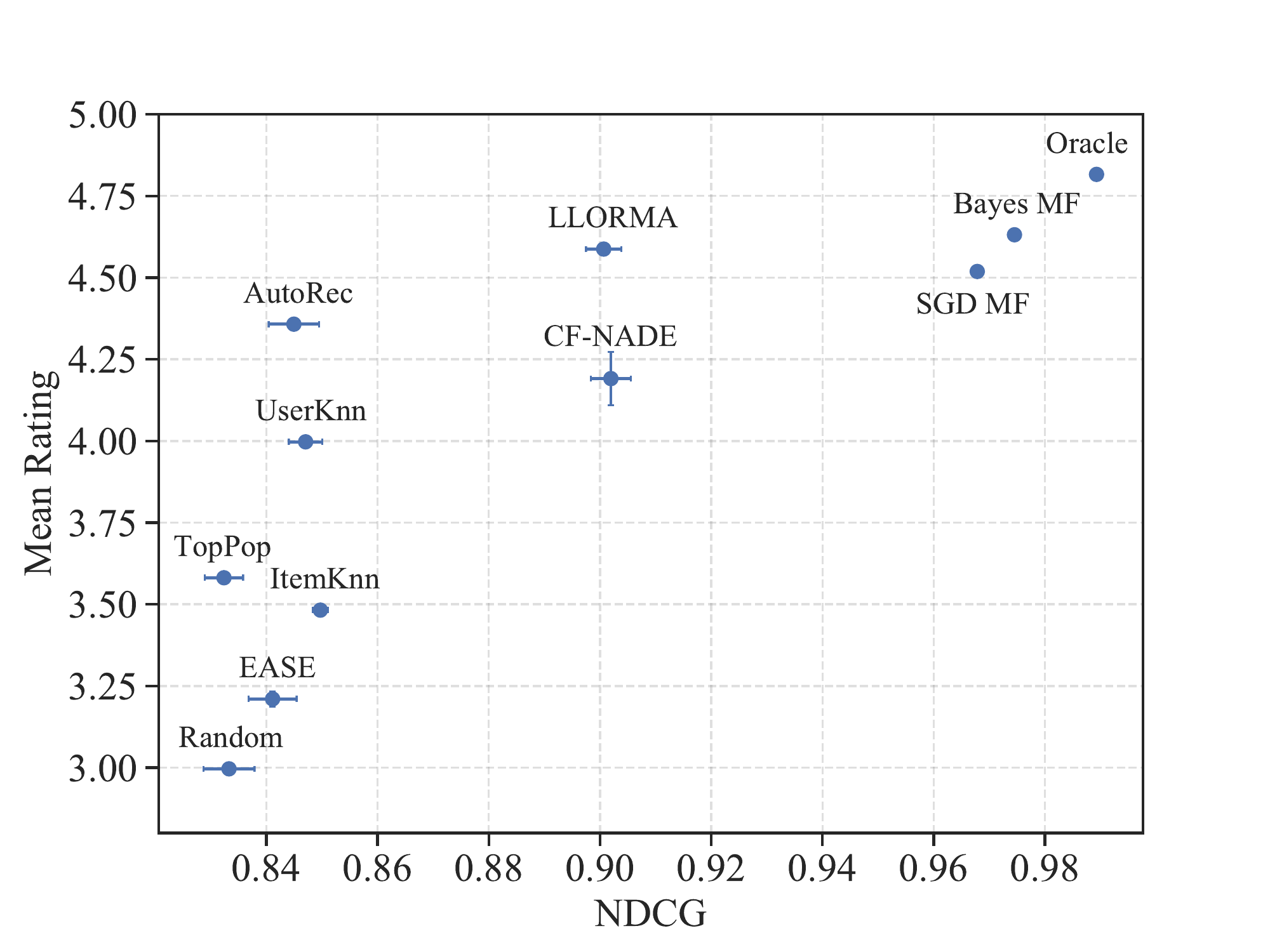}}
    \subfloat{\includegraphics[width=0.45\textwidth]{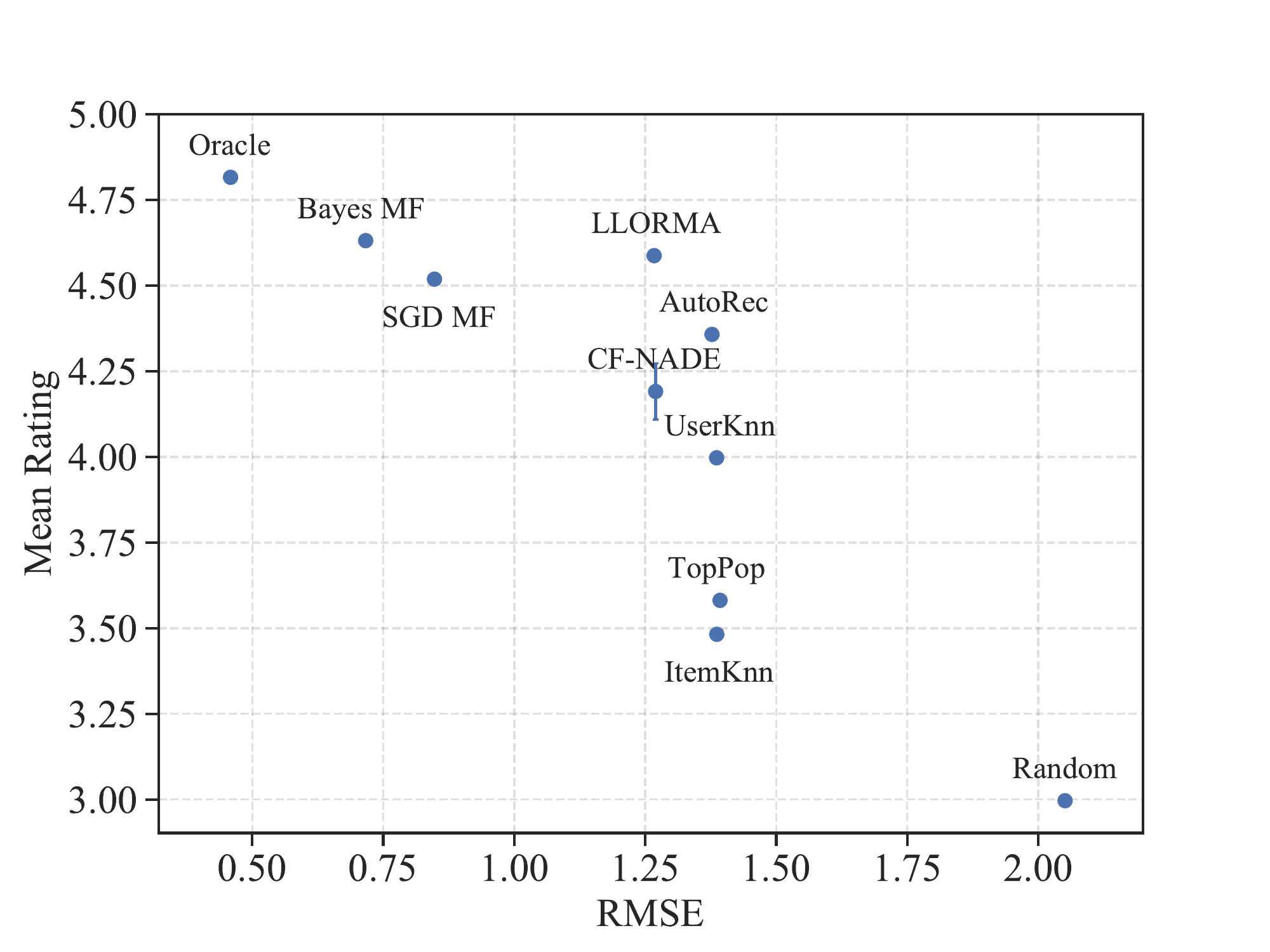}}
    \caption{Left: The nDCG@20 plotted against the mean user ratings of all recommended items on the \texttt{topics-dynamic} environment. Right: The RMSE plotted against the mean user ratings. nDCG and RMSE are averaged across 5 folds on the offline dataset associated with the environment, user ratings are averaged across 10 trials. Each point represents a single model evaluation with error bars representing 95\% confidence intervals.}
    \label{fig:main-plot}
\end{figure*}

We first show that offline metrics can act as a good proxy for online performance by replicating the offline-first evaluation methodology in a controlled setting. We compute RMSE and nDCG on an offline dataset for each recommender, simulate the interactive process of recommendation, compute the average user ratings of recommended items, and then compare the offline and online metrics. Figure~\ref{fig:main-plot} shows such a comparison on the \texttt{topics-dynamic} environment. While there is a strong correlation between nDCG and mean user rating, we note that improvements in nDCG past a certain point suffer from diminishing improvement in mean user ratings. We examine these effects in more details in Section~\ref{sec:offline-online-expts}.

We then consider low-data regimes, in which recommenders do not have access to a large offline dataset (Section~\ref{sec:exploration}).
In this case, augmenting existing recommenders with exploration techniques leads to better prediction accuracy over the \emph{full population} of users and items.
However, we show that the performance benefit of such augmentations,  
as measured by the average user rating of the recommended items, varies depending on the recommender algorithm and the underlying environment. 
Finally, we look at the relationship between a recommender's item coverage and online performance. 
We observe that the correlation between the two metrics depends on the environment, indicating that further investigation is necessary.

Taken together, our large-scale simulation results suggest that offline metrics can be a useful tool when online evaluation is not possible. 
However, they also bring into question the value of chasing small improvements in predictive performance, since those results might only lead to very small improvements in the online setting. This is especially true when data is plentiful and state-of-the-art recommendation algorithms can predict near-optimally.
Instead, researchers should focus on a holistic evaluation of their algorithms; taking into account metrics that look beyond predictive accuracy and considering issues of measurement and sampling in the construction of the datasets they use.

Since there are many interesting research questions that can be studied through simulation, we open-source our simulation package. Our package is designed with large-scale evaluation in mind, and reproduces a number of popular and state-of-the-art recommenders.
We also make available many environments, while making it easy to extend the package with new environments.
It is our hope that this package can be integrated into recommender algorithm development and evaluations.
\section{Related Work}

\paragraph{Other simulations}
There has recently been increased interest in studying recommenders through simulation. \citet{chaney2018algorithmic} propose an environment in which users have limited knowledge of their utility, and show that recommendation algorithms homogenize user behavior within their simulations. \citet{schmit2018interaction} evaluate the performance of ridge regression and matrix factorization in a two timestep simulated dynamical setting. \citet{ie2019recsim} and \citet{rohde2018recogym} both propose simulation frameworks that are focused on evaluating reinforcement-learning based recommenders. \citet{mansoury2020feedback} and \citet{jiang2019degenerate} use simulation to study the negative effects of feedback loops in recommender systems.
Our work distinguishes itself from prior simulation studies by being the first to investigate a wide range of recommendation algorithms across a large number of environments. Furthermore, we are the first to investigate simulations at the scale of common benchmark datasets, while still running for many timesteps.

\paragraph{Online recommendation}
When in production, recommender systems interact iteratively with a changing environment where the set of users and items is not constant and user preferences evolve alongside the recommender.
Several empirical studies on deployed recommender systems identify inconsistencies in online and offline performance~\cite{beel2015comparison,mogenet2019predicting,rossetti2016contrasting}, 
while others show how richer sets of offline metrics can be used to predict online performance~\cite{maksai2015predicting}. Our work brings a systematic lens to this problem to understand more broadly the validity of practices around algorithm design and evaluation.

Many recommendation algorithms have been designed to address the difficulty of online recommendation in dynamical environments.
Classically, time-aware models exploit the sequential nature of recommendation by incorporating temporal context \cite{koren2009collaborative, vinagre2015overview, campos2014survey}. More recently, a body of work treats online recommendation as a causal inference problem where the recommender model must de-bias logged training data \cite{schnabel2016recommendations, matuszyk2015forgetting, saito2020unbiased, sinha2016deconvolving}. Lastly, others seek to improve recommender systems by directly addressing the online problem either through exploration strategies or reinforcement learning algorithms \cite{li2010contextual,kawale2015efficient, ie2019reinforcement, chen2019top}.

\paragraph{Alternative metrics}
Our work complements research on the societal effects of recommender systems, which considers alternative metrics including diversity, utility, serendipity and fairness \citep{adomavicius2013, nguyen2014exploring, fleder2009blockbuster, singh2018fairness, yao2017fairness}. Many authors have examined the limitations of accuracy as the sole metric for evaluating recommenders and have sought to define alternate metrics. \citet{herlocker} proposed a variety of metrics for assessing a recommender's coverage, diversity, novelty, and serendipity, while \citet{kaminskaseval} provide a more recent survey of the approaches for training recommenders with respect to these alternative metrics.
Recently, \citet{dean2020recommendations} introduced a measure of reachability which combines ideas of coverage with user agency in interactive systems.

\section{Reproduced Recommenders}

We evaluate eleven recommenders including baseline models, neighborhood models, factorization machine models, and several recent deep models. We choose to investigate recommenders mentioned by \citet{rendle2019difficulty} as these models have all been run on the MovieLens10M dataset, giving us a starting point of comparison. The recommendation algorithms that we consider choose items to recommend based on a  \emph{relevance prediction}.
Except when specified, recommendations are \emph{greedy}, meaning that the item with the highest relevance prediction will be recommended.
Therefore, the main difference between the following algorithms is their prediction component. All the models we reproduce only make use of ratings when making predictions, in future work we wish to also evaluated content-based and temporal models.
We focus on the settings of rating prediction, where models are tuned so that relevance predictions match ratings (e.g. RMSE).



\begin{itemize}
    \item \texttt{TopPop} - The TopPop algorithm recommends the most popular items to every user without personalization. The popularity of each item is measured by its average rating.
    \item \texttt{ItemKNN} - The ItemKNN algorithm is a collaborative filtering method using $k$-nearest neighborhood and item similarities \cite{wang2006unifying}. We implement ItemKNN in the same way as \citet{Surprise}. 
    \item \texttt{UserKnn} - The UserKnn algorithm is identical to ItemKnn, except that it uses user features instead of item features \cite{wang2006unifying}.
    \item \texttt{Oracle} - The oracle recommender has access to the internals of the simulated environment, and will recommend the item with the highest true rating at each time step. Since this oracle baseline is still greedy, it does not plan for environment dynamics. Additionally, since actual ratings are generated with some noise, the RMSE of the oracle baseline is not zero.
    \item \texttt{Random} - This baseline predicts ratings uniformly at random.
    \item \texttt{SGD MF} - A factorization machine implemented in LibFM \cite{rendle:tist2012}. The model is trained using SGD. \item \texttt{Bayes MF} - Another variant of factorization machines implemented in LibFM. The model is trained and the hyperparameters are automatically tuned using MCMC.
    \item \texttt{AutoRec} - An autoencoder framework for collaborative filtering \cite{sedhain2015autorec}. We train the algorithm with RMSProp and use the item-based version \texttt{I-AutoRec}. Our implementation makes use of the source code provided by the authors.\footnote{\url{https://github.com/mesuvash/NNRec}}
    \item \texttt{CF-NADE} - A neural autoregressive architecture for collaborative filtering \cite{zheng2016neural} trained with Adam. We adapt a publicly available implementation for our experiments.\footnote{\url{https://github.com/JoonyoungYi/CFNADE-keras}}
    \item \texttt{LLORMA} - LLORMA \cite{lee2016llorma} is a generalization of low rank matrix factorization techniques. LLORMA approximates the rating matrix as a weighted sum of low-rank matrices. We adapt a publicly available implementation for our experiments.\footnote{\url{https://github.com/JoonyoungYi/LLORMA-tensorflow}}
    \item \texttt{EASE} - EASE \cite{steck2019embarrassingly} is a linear model designed for sparse data, especially implicit feedback data in recommenders. We do not include this recommender when computing RMSE as it outputs non-normalized relevance scores.
\end{itemize}

\section{The \reclab Simulated Environments}
\label{sec:environments}

In this section we summarize the environments that we provide through our simulation framework. We consider both environments where users must consume the single item  that is recommended to them and environments where users can choose from a slate of items. Unless mentioned otherwise we set our environments to have $1000$ users and $1700$ items, which is similar to the MovieLens 100K dataset. All the environment hyperparameters values we used are available in the experiments directory of the provided code.\footnote{Experiment code for the paper can be downloaded at https://github.com/berkeley-reclab/RecLab}

\paragraph{\texttt{topics-static}}
In the \texttt{topics-static} environment, each item is assigned to one of $K$ topics and users prefer certain topics, this is similar to the simulation presented by \citet{ie2019reinforcement}. The preference of user $u$ for items $i$ of topic $k_i$ is initialized as $\pi(u,k_i) \sim Unif(0.5, 5.5)$, while the topic $k$ of item $i$ is chosen randomly from the set of all topics. When user $u$ is recommended item $i$ it will rate the item as $r_t(u,i) = \clip(\pi(u,k_i) + \epsilon)$
where $\epsilon \sim \mathcal{N}(0, \sigma^2)$ represents exogenous noise not modeled by the simulation, and $clip$ truncates values to be between $1$ and $5$.

\paragraph{\texttt{topics-dynamic}}
In \texttt{topics-dynamic} items are rated in the same way as \texttt{topics-static}. In this setting however, user preferences can change as a result of the items they consume. Since past work has shown that users might become more interested in a topic through repeated exposure \cite{ge2020understanding, mansoury2020feedback}, we incorporate this phenomenon into our model. If item $i$ is recommended to user $u$ then their preferences are updated as
\begin{align*}
    \pi_{t+1}(u, k)
     &\leftarrow \clip(\pi_t(u, k) + a)\quad k = k_i,\\
    \pi_{t+1}(u, k') &\leftarrow \clip\left(\pi_t(u, k') - \frac{a}{K - 1}\right) \quad \forall k' \neq k_i,
\end{align*}
where $a$ is a fixed affinity parameter. 
Another well-studied phenomenon is the fact that users get bored from being recommended the same topic in a short period of time \cite{kapoor2015just, warlop2018fighting}. We model this as:
\begin{align*}
r_t(u,i) = \clip(\pi_t(u,k_i) - \lambda\mathbf{1}\{&\text{topic}~k_i~\text{observed}\geq \tau~\text{times within}\\
&~m~\text{previous timesteps}\} + \epsilon).
\end{align*}
The effect of boredom arises from 
three parameters: memory length $m$, boredom threshold $\tau$, and boredom penalty $\lambda$. If a user observes the same topic more than $\tau$ times within the last $m$ timesteps then their ratings is penalized by $\lambda$. 


\paragraph{\texttt{latent-static}}
In the \texttt{latent-static} environment, we represent users and items with $d$-dimensional latent feature vectors. This is a common assumption when developing factor models \cite{koren2008factorization, bell2007modeling, koren2009matrix}, and allows us to investigate a different user-item representation than the topics-based simulations. The rating of user $u$ on item $i$ is computed using these latent vectors as well as bias terms: $r(u,i) = \clip(\mu_0 + c_u + b_i + \mathbf p_u^\top \mathbf q_i + \epsilon)$,  
where $\mu_0 = 3$ is a global bias, $c_u \sim \mathcal{N}(0, 0.25)$ is the bias of user $u$, $b_i \sim \mathcal{N}(0, 0.25)$ is the bias of item $i$, $\mathbf p_u \sim \mathcal{N}(0, \sqrt{0.5 / d})$ is the latent factor of user $u$, $\mathbf q_i \sim \mathcal{N}(0, \sqrt{0.5 / d})$ is the latent factor of item $i$, and $\epsilon \sim \mathcal{N}(0, \sigma^2)$. 
\texttt{RecLab} also includes a \texttt{latent-dynamic} environment with a similar concept of boredom and affinity change as \texttt{topics-dynamic}. However, none of our experimental results use \texttt{latent-dynamic}.

\paragraph{\texttt{ML-100K}}
The \texttt{ML-100K} environment is identical to the\\ \texttt{latent-static} environment, except that the parameters are generated based on the MovieLens 100K (ML\_100K) dataset \cite{harper2015movielens}. We train a LibFM factorization model on the ML\_100K dataset, with hyperparameters tuned to achieve low RMSE through cross validation. We then extract the model's biases and latent factors to initialize the environment. We use \texttt{ML-100K} to confirm that our experiments generalize to situations where the simulator's parameters are initialized using real user interaction data.

\paragraph{\texttt{latent-score}}
The \texttt{latent-score} environment was first proposed in Section~5.2.3 of \citet{schmit2018interaction}. The difference between \texttt{latent-score} and \texttt{latent-static} is that users have partial knowledge of an item's value. They use this partial information, along with the recommender's predicted score, to select an item from a slate of recommended items. We evaluate this environment with $170$ users and $100$ items due to computational limitations.


\paragraph{\texttt{beta-rank}} 
This environment was introduced by \citet{chaney2018algorithmic}. It is similar to \texttt{latent-score}: users know part of the value for each item and users/items are represented by latent vectors. In \texttt{beta-rank} the rating of a user $i$ on an item $j$ is given by $r(u, i) \sim Beta(\mathbf p_u^\top \mathbf q_i, \sigma^2)$, 
where $\mathbf p_u$ is the latent vector for user $i$, $\mathbf q_i$ is the latent vector for item $i$, and the Beta distribution is parametrized according to its mean and variance. In this setting users chose from a slate of items based upon their observed utility and the recommender's ranking. We evaluate this environment with $170$ users and $100$ items due to computational limitations.


\section{The Relationship Between Offline Metrics and Online Performance} \label{sec:offline-online-expts}

\begin{figure*}[ht]
    \centering
    \includegraphics[width=0.8\textwidth]{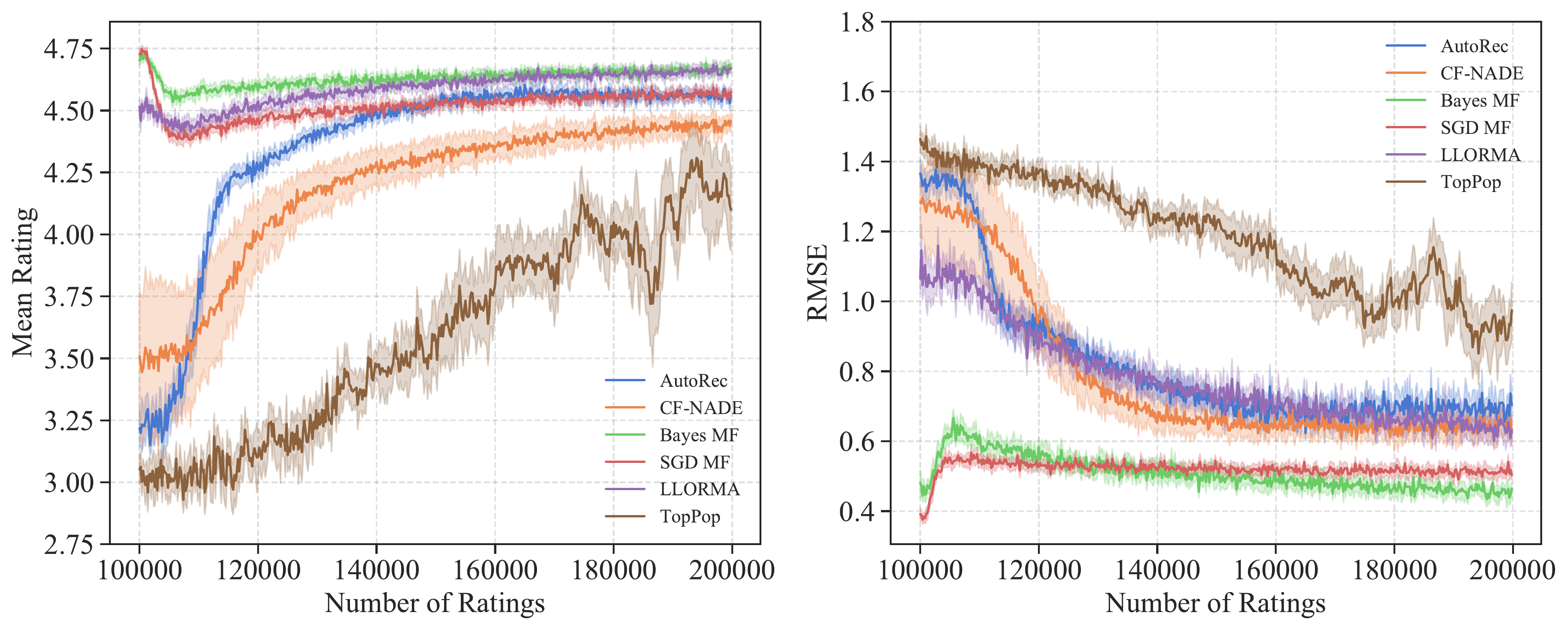}
    \caption{The performance of select recommenders over time on the \texttt{topics-dynamic} environment. The left plot shows the mean rating of the items that are recommended at each timestep. The right plot shows the RMSE between the predicted and true ratings of the recommended items at each timestep. Both plots are created by averaging over $10$ experiment trials. The shaded areas represent 95\% confidence intervals.}
    \label{fig:topics-dynamic-timeseries}
\end{figure*}
In this section we explore the relationship between offline metrics and online performance across all the environments described in Section~\ref{sec:environments}. Given an environment, we evaluate each recommender by following these steps:
\begin{enumerate}
    \item We create an offline dataset by sampling user-item pairs without replacement and get their ratings from the environment. We sample pairs uniformly, which removes the sampling bias introduced by data collected when a recommendation algorithm is already deployed.
    This allows us to focus on the effects of environment and recommender dynamics.
    \item We tune the recommenders on the offline dataset using grid search on the hyperparameters. Our search aims to minimize the mean RMSE evaluated using 5-fold cross-validation.
    We use these hyperparameters throughout the experiment.
    \item We begin by training each recommender using our offline dataset. At each timestep, online users are uniformly sampled from the set of all users. We recommend items to the sampled users and observe their ratings.
    Lastly, we retrain the recommenders on all the acquired data so far, and repeat until the end of the experiment.
    We repeat for $10$ trials for every recommender on each environment.
\end{enumerate}
This method of experimentation closely mimics the offline-first evaluation method mentioned in Section~\ref{sec:intro}.
Notice that we control for factors which may impact online performance, including the initial sampling distribution and the distribution of online users, so that our results illustrate exclusively the effect of environment and recommender dynamics.
Furthermore, we consider simple online deployment, retraining from scratch whenever data is added and using the same hyperparameters throughout.
While it is possible to improve these methods, for example by performing just a few gradient steps, we wish to avoid any confounding factors whose effect on performance aren't well understood.

Throughout this section, we consider greedy recommendation: we always recommend the item with the highest predicted relevance that hasn't already been recommended to each user .

\subsection{Dynamic Environment}\label{sec:topics-dynamic}
In this section we discuss results on \texttt{topics-dynamic}, an environment where user's preferences dynamically change over time. We sample 100k initial ratings on which to tune the recommenders, at every timestep we sample $200$ users to recommend items to, and we run the simulation until we observe 200k ratings.
As shown in Figure~\ref{fig:main-plot}, nDCG@20\footnote{We tried many different values for nDCG@k and got near identical behavior for all $k$.} and RMSE are predictive of mean user rating. While it is well known that no offline metric can perfectly track online performance, the reason for these discrepancies has not been extensively studied.

We explain some of these discrepancies by exploring the full timeseries behavior of \texttt{topics-dynamic} for a select group of recommenders. The left plot in Figure~\ref{fig:topics-dynamic-timeseries} shows the average rating over the recommended items at each timestep, while the right plot shows the RMSE  between the predicted and true ratings of the recommended items at each timestep. We emphasize that the RMSE shown in Figure~\ref{fig:topics-dynamic-timeseries} is only computed with respect to the recommended items at each timestep, and hence is not a measure of  accuracy on the whole population of ratings. We are particularly interested in reasons why two recommenders might have:  similar offline nDCG/RMSE but dissimilar mean user rating, similar mean user rating but dissimilar offline nDCG/RMSE, or dissimilar nDCG and RMSE. With these goals in mind we make the following observations: 
\begin{itemize}
    \item \textbf{Recommenders can affect user preferences in ways not captured by offline metrics.} \texttt{TopPop} improves its performance significantly as time progresses. There is a large difference in mean rating between it and \texttt{Random}, despite the fact that both recommenders have very close nDCG. As we discuss in Section~\ref{sec:topics-static}, \texttt{TopPop}'s improvements are due to the environment's dynamics. 
    \item \textbf{The size of the initial dataset affects the offline ranking of recommenders.} \texttt{LLORMA} performs well starting at the first timestep. This seems to contradict its relatively low NDCG on the offline dataset. However, further investigation reveals that LLORMA's NDCG increases to 0.948 when trained with $100,000$ datapoints instead of the $80,000$ available during 5-fold cross-validation.
    \item \textbf{Recommender dynamics can affect their performance in ways not captured by offline metrics.} \texttt{AutoRec} performs much better than its offline nDCG and RMSE indicate. Unlike \texttt{LLORMA}, the initial performance of \texttt{AutoRec} is bad. At the first timestep, it barely performs better than \texttt{Random}.
    However, once it observes data that is sampled from its own recommendations, it is able to quickly match the performance of the best recommenders.
    \item \textbf{The same user dynamics can have a positive effect on one recommender, while simultaneously having a negative effect on another.} \texttt{CF-NADE} improves much more slowly than \texttt{LLORMA}, explaining the difference in rating between both algorithms despite the similar nDCG.\footnote{This observation also holds for \texttt{UserKnn} and \texttt{ItemKnn} although we do not show all these recommenders in Figure~\ref{fig:topics-dynamic-timeseries} to reduce clutter.} As we show in Section~\ref{sec:topics-static}, this is primarily due to the negative effect of the user dynamics on \texttt{CF-NADE}.
    \item \textbf{A low RMSE is sufficient for selecting high-value items, but it is not necessary.} \texttt{SGD MF} and \texttt{Bayes MF} both perform on-par with \texttt{LLORMA}, despite the fact that \texttt{LLORMA} has a worse offline RMSE and per-timestep RMSE. This is because 
    RMSE captures a recommender's ability to predict ratings, whereas mean user rating captures a recommender's ability to identify high-value items.
\end{itemize}

\subsection{Static Environment}\label{sec:topics-static}
In this section, we investigate the performance of recommenders on the \texttt{topics-static} environment. We focus on comparing these results with those obtained on \texttt{topics-dynamic} to disentangle phenomena caused by the environment dynamics and those caused by the recommender dynamics. We initialized \texttt{topics-static} with the same user preferences and item topics as \texttt{topics-dynamic}. Furthermore, we ensure the offline dataset is the same for both environments. As a result each recommender's hyperparameters are the same as in the \texttt{topics-dynamic} setting.

\begin{figure}[h]
    \centering
    \includegraphics[width=0.45\textwidth]{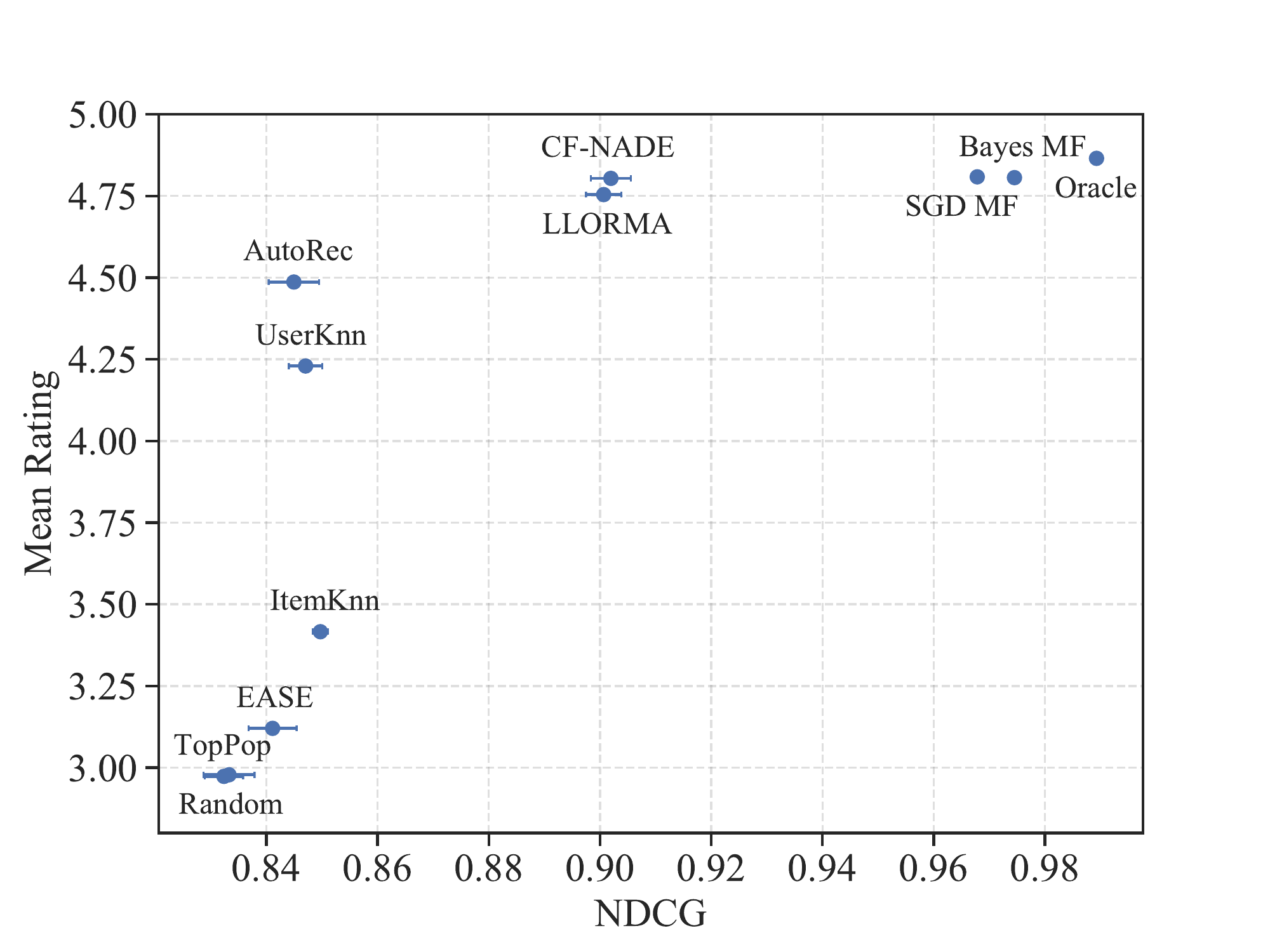}
    \includegraphics[width=0.45\textwidth]{topics-static-rmse}
    \caption{Left: The nDCG@20 plotted against the mean user ratings of all recommended items on the \texttt{topics-static} environment. Right: The RMSE plotted against the mean user ratings. NDCG and RMSE are averaged across 5 folds on the offline dataset associated with the environment, user ratings are averaged across 10 trials. Each point represents a single model with error bars representing 95\% confidence intervals.}
    \label{fig:topics-static}
\end{figure}

Figure~\ref{fig:topics-static} compares the nDCG@20 of each recommender on the offline dataset with the average user rating across all timesteps. In this setting, nDCG is still positively correlated with mean rating, and while most of the observations made in Section~\ref{sec:topics-dynamic} are still a concern even when there are no environment dynamics at play, we identify two notable differences.
First, \texttt{TopPop} performs on par with \texttt{Random}. Without environment dynamics, the underlying preferences remain uniformly distributed, so popularity is not predictive.
This is in contrast to \texttt{topics-dynamic} where the average preference for each topic is also initialized to be roughly $3$, but by the end of the trial with \texttt{TopPop}, two of the topics have average affinities higher than $4.5$. \texttt{TopPop} pushes user preferences toward certain topics, inducing item popularity effects in the data. 
Furthermore, \texttt{CF-NADE} performs significantly better, showing that the same user dynamics can have positive or negative effects on performance depending on the recommendation algorithm. This emphasizes the importance of evaluating general-purpose recommenders across a diverse range of datasets and environments. We observed similar results when comparing RMSE to mean rating in the \texttt{topics-static} environment.

\subsection{Other Environments}
\begin{figure}[h]
    \centering
    \subfloat{\includegraphics[width=0.5\textwidth]{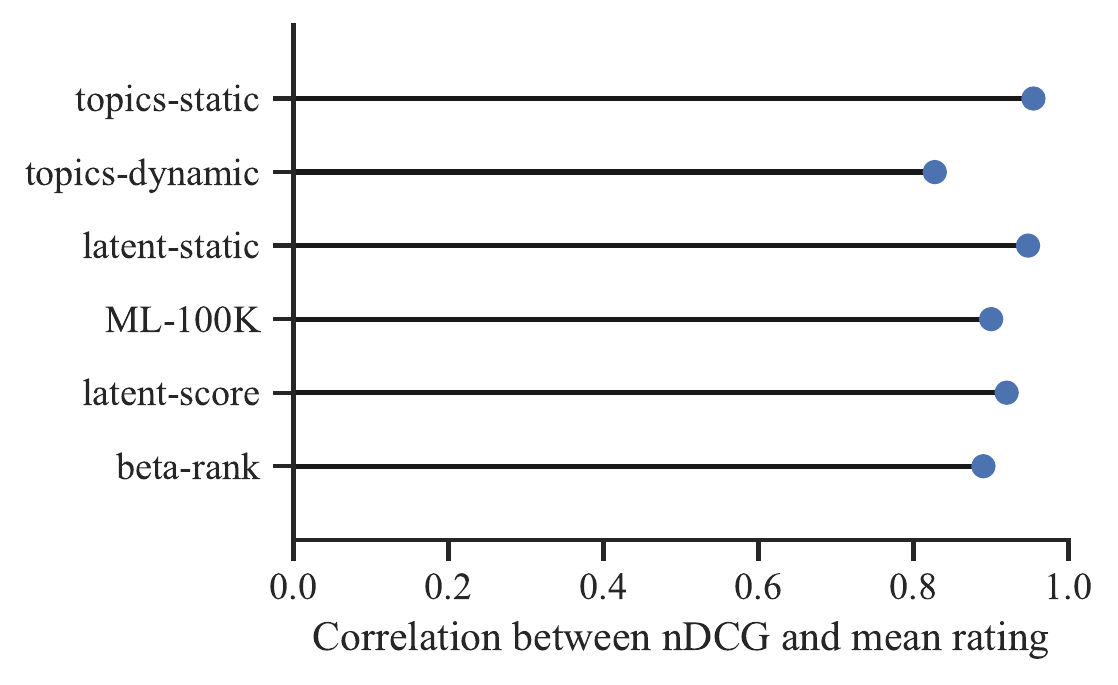}}
    \subfloat{\includegraphics[width=0.5\textwidth]{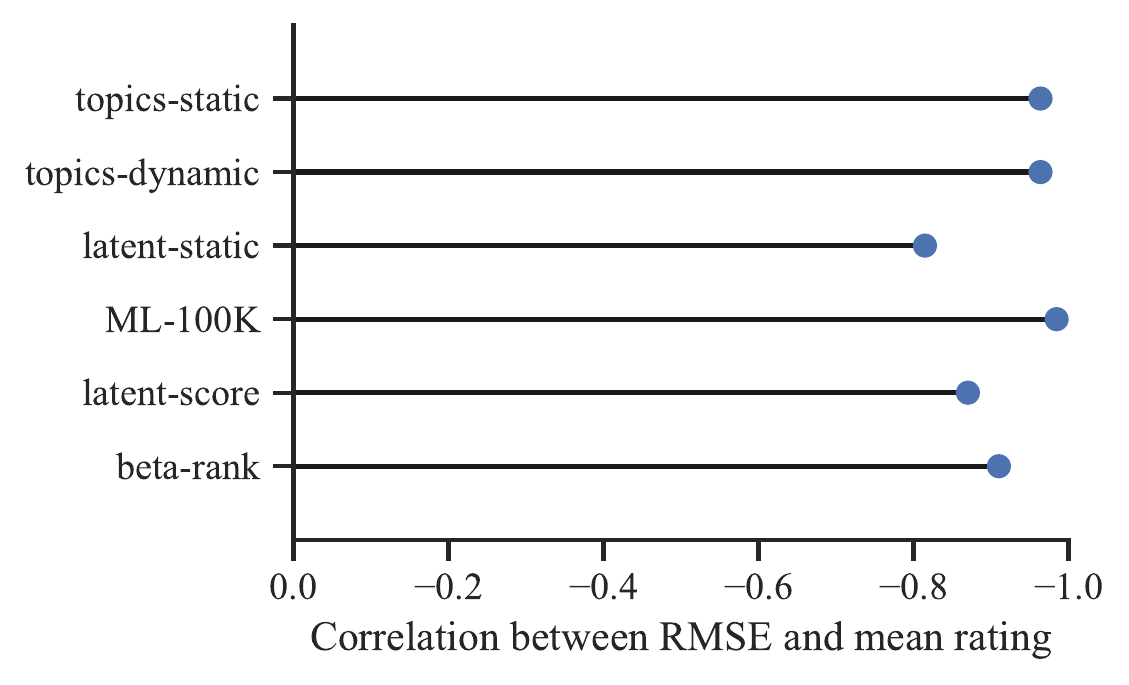}}
    \caption{Left: The Spearman correlation between the nDCG@20 and the mean user ratings of all recommended items across all environments. Right: The Spearman correlation between the RMSE and the mean user ratings of all recommended items across all environments.}
    \label{fig:all-envs}
\end{figure}

So far we have only demonstrated that RMSE and nDCG are predictive of online performance for two environments: \texttt{topics-static} and \texttt{topics-dynamic}. To ensure that this observation is robust, we benchmark the recommendation models across all environments from Section~\ref{sec:environments}.  The left plot of Figure~\ref{fig:all-envs} shows the Spearman rank correlations \cite{spearman} between the nDCG@20 and the mean rating, while the right plot shows the correlations between the RMSE and the mean rating. We see that both nDCG and RMSE are predictive of online performance across all environments. For \texttt{latent-score} and \texttt{beta-rank} we sample 1000 initial ratings and run the simulation until we have 2000 ratings. For all other environments, we sample 100k ratings and run the simulation until we have 200k ratings. 
\section{The Effects of Exploration}
\label{sec:exploration}

\begin{figure*}[ht]
    \centering
    \includegraphics[width=0.45\textwidth]{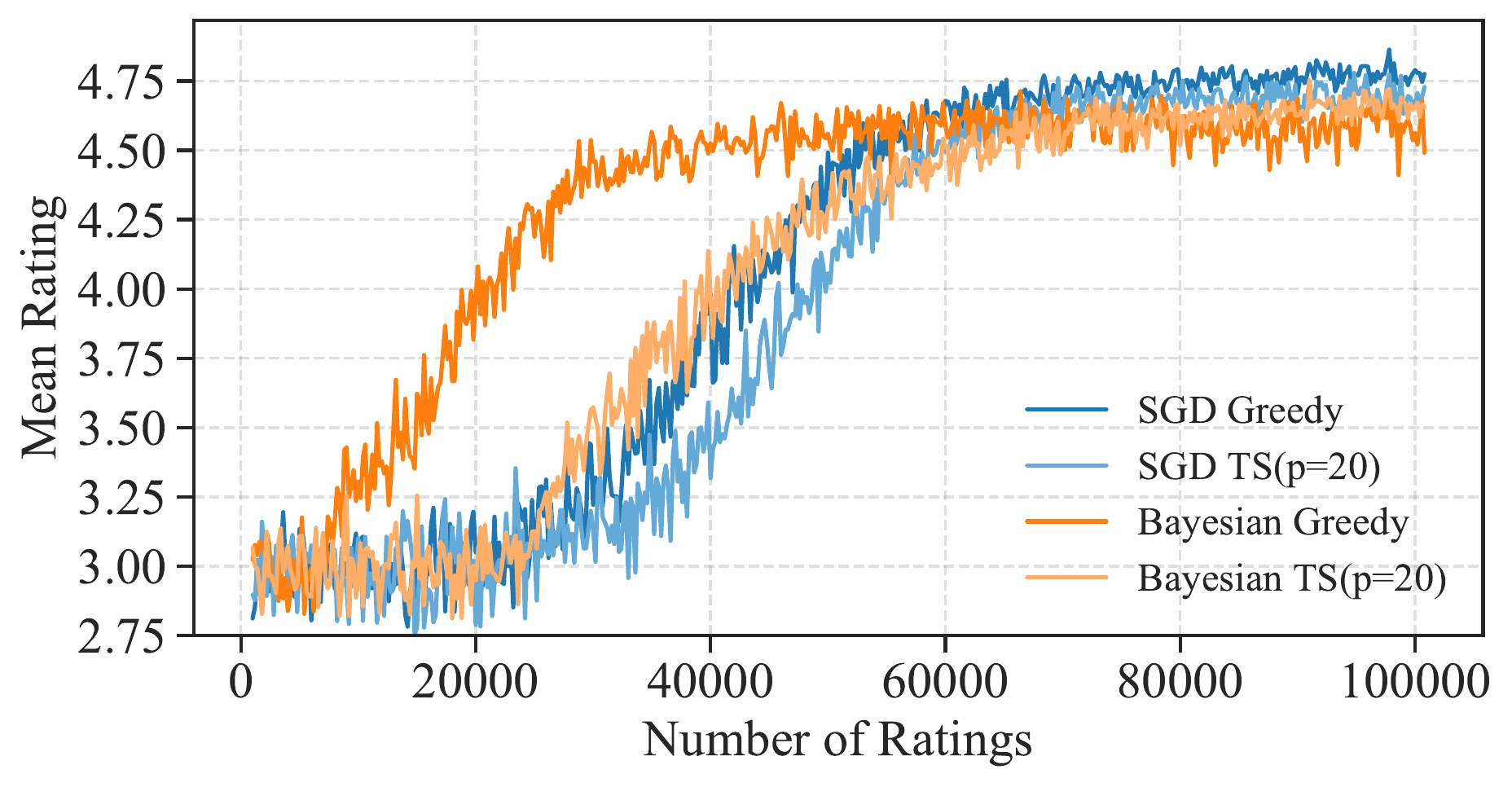}
    \includegraphics[width=0.21\textwidth]{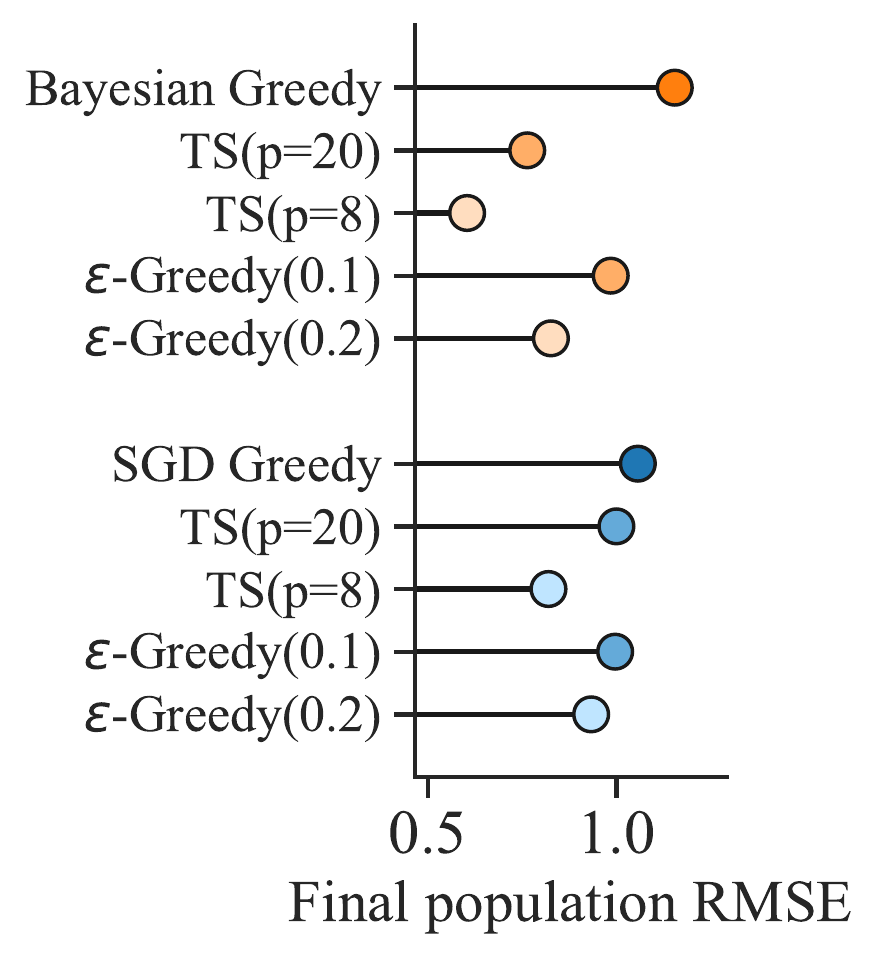}
    \includegraphics[width=0.31\textwidth]{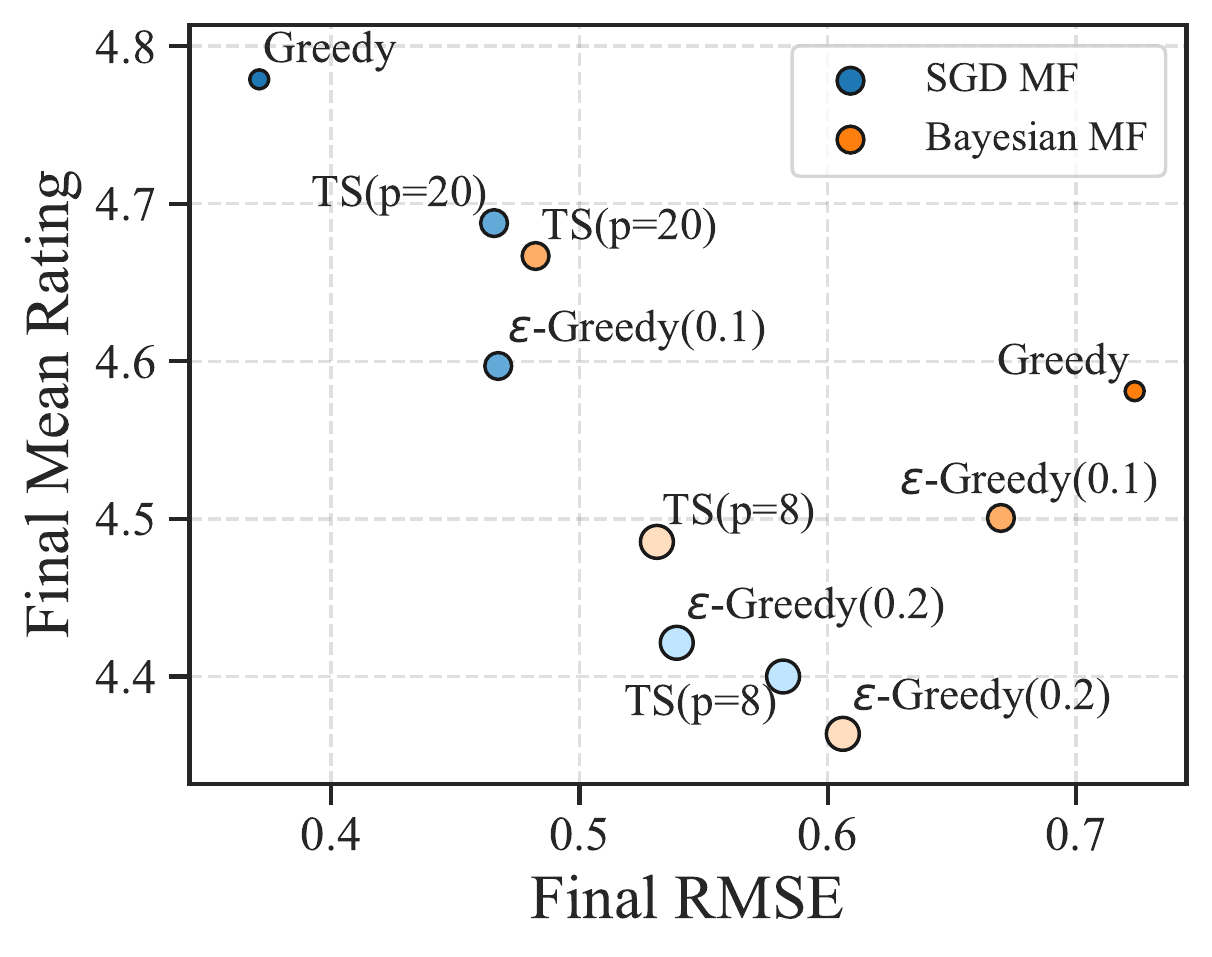}
    \caption{Exploration strategies on \texttt{topics-static-lowdata}. Left: The mean rating starts low and increases as the recommenders get more data. Middle: The population RMSE at the end of the experiment measures the overall identification. Right: The final RMSE and mean rating indicate the online performance. Final metrics are computed as the average of the 1,000 final ratings.}
    \label{fig:explore_topics}
\end{figure*}

Previously we considered a controlled and high-data setting. We now consider the challenges posed by a low-data setting and whether these challenges can be mitigated by exploration. In particular, we investigate whether simple modifications to recommenders designed and tuned in offline settings can further improve their performance in online settings.
The results so far consider the repeated application of a greedy selection rule.
Common wisdom argues that a purely exploitative strategy will be suboptimal in a sequential decision making setting.

We therefore consider probabilistic selection rules, which have been suggested by the literature on bandit strategies in recommendation systems \cite{kawale2015efficient}. Prior work suggests that employing non-deterministic selection rules improves regret bounds over the horizon of recommendations \cite{li2010contextual}.
We specifically consider two widely used non-deterministic selection strategies:
\begin{enumerate}
    \item \textbf{$\epsilon$-Greedy}: The probability of choosing item $i$ for user $u$ among a set of available items $\mathcal{I}_u$ is:
    \begin{align*}
        \mathbb{P}\{i\} =\begin{cases} 1 - \epsilon & \text{if } i = \arg\max_{i\in \mathcal{I}_u} \hat r(u, i) \\
        \frac{\epsilon}{|\mathcal{I}_u|} & \text{otherwise.}
        \end{cases}
    \end{align*}

    \item \textbf{Thompson Sampling}: The probability of choosing an item is proportional to an increasing function $\phi(\cdot)$ of the predicted ratings:
    \begin{equation*}
      \mathbb{P}\{i\} \sim \phi(\hat r(u,i)).
    \end{equation*}
    We use the function $\phi(r) = r^p$, where $p$ is a parameter which controls the spread of the sampling distribution.
\end{enumerate}

To measure the effects of exploration, we focus on a low data setting, where recommenders initially have access to one rating per user on average.
We consider the \texttt{topics-static-lowdata} environment, a version of \texttt{topics-static} in which 1,000 randomly selected ratings are initially revealed.
We consider two of the highest performing  algorithms: \texttt{\texttt{SGD MF}} and \texttt{Bayes MF}.
The recommenders are augmented with the stochastic selection rules described above.
We consider four settings of exploration: $\epsilon$-Greedy with $\epsilon=0.1$ and $\epsilon=0.2$ and Thompson Sampling (TS) with $p=20$ (less spread) and  $p=8$ (more spread).
We follow the experimental procedure outlined in Section~\ref{sec:offline-online-expts}, with \texttt{num\_trials=1}.
Since the initial dataset is too small for tuning, each recommender uses the same hyperparameter tuning as for our \texttt{topics-static} experiments.

\subsection{Performance Effects}

We first consider the performance on~\texttt{topics-static-lowdata}.
The leftmost panel in Figure~\ref{fig:explore_topics} plots the average rating of recommended items over time, 
demonstrating how the performance evolves for various recommendation strategies.
It is immediately notable that greedy \texttt{Bayes MF} achieves a worse final performance than greedy \texttt{\texttt{SGD MF}}, despite its superior performance on full size environments and its more rapid initial increase.
Additionally, while Thompson sampling improves the performance of \texttt{Bayes MF},  we do not see the same benefit for \texttt{SGD MF}.
The rightmost panel in
Figure~\ref{fig:explore_topics} summarizes this finding: the vertical position of each point indicates the final mean rating, which is computed as the average of the 1,000 final ratings.
Comparing the eight recommendation strategies,
for \texttt{SGD MF}, exploration strategies with higher amounts of randomness achieve lower performance.

To further understand the quality of the recommenders, we examine the RMSE of the recommended items, as well as the \textit{population} RMSE,
which serves as a measure of overall prediction accuracy by considering every user and item pair in the system.
The middle panel in Figure~\ref{fig:explore_topics} shows the final population RMSE.
Evaluated by this metric, all exploration strategies have a positive effect, with more randomness leading to a larger decrease in population RMSE.
While useful for understanding the effects of exploration strategies, this measure is unrealistic, since deployed systems can only observe the ratings of {recommended} items.
The right panel in Figure~\ref{fig:explore_topics} plots the \emph{observed} final RMSE along the horizontal axis, which is computed over the final 1,000 ratings.
For \texttt{SGD MF}, greedy performs the best, while exploration strategies with less randomness perform better than those with more.
For \texttt{Bayes MF}, the pattern is reversed.

These experiments lend insight to two important issues: the performance of algorithms in low data settings, and how properties of the underlying rating distribution influence the effect of biased sampling.
Even though \texttt{Bayes MF} was the highest performing algorithm when it began with access to 100,000 ratings, its performance suffers in this low data setting.
The difference is likely related to the biased distribution that arises from online recommendations. \texttt{Bayes MF} integrates
hyperparameter tuning into its inference process, and therefore the regularization values are automatically determined. On the other hand, we do not re-tune hyperparameters for \texttt{SGD MF} online, which might lead to additional exploration and more robustness towards the bias in the online data.
Furthermore, we ran an expanded set of exploration-based experiments on the \texttt{latent-static} environment. We found that the observed differences between results in \texttt{topics-static} and \texttt{latent-static} indicate that sampling biases affect performance in a way that depends on the underlying rating distribution.
These observations highlight the fact that online deployment of recommender algorithms comes with setting-dependent caveats and challenges.

\section{Discussion}
Our experiments led to encouraging results regarding the way recommender systems are currently evaluated. In performing these evaluations, we surfaced observations that  paint a more multifaceted view of a recommender's performance. In this section, we touch on three major areas we hope future algorithm developers will keep in mind when designing new recommender systems.

\subsection{The Role of Simulation in Evaluation}
Our work makes heavy use of simulation to investigate recommender systems. It is worthwhile to view such simulations with a critical eye, as they have clear limitations.
We justify each design choice in our environments by referring to data-backed prior work, and we also run experiments on existing environments that are vetted by the research community. However, none of these environments are a perfect recreation of the real world, which would involve the immensely complex task of reproducing human behavior. Although this is a limitation of our work, there is significant value in studying algorithms and metrics in a simplified setting. The use of simulation to study simplified settings is more accessible and thus reproducible.
It also lower the stakes, leading to faster iteration when designing new algorithms.
Crucially, the ability to control for many factors allows for a more mechanistic understanding of observed phenomena.
This is a widely accepted fact in theory-oriented work \cite{chu2011contextual, wang2013theoretical}, and is starting to be adopted within more empirical settings. For example, simulation has been widely accepted within the field of reinforcement learning as a tool to benchmark algorithms. These simulations vary from semi-realistic physical models \cite{todorov2012mujoco} to video-games with little grounding in reality \cite{mnih2013playing, berner2019dota}. Furthermore, the fields of computer vision and natural language processing have also started to assess their widescale evaluation practices by experimenting on simplified tasks that act as a proxy to real-world tasks \cite{recht2019imagenet, miller2020effect}.

A recommender system or metric performing well in simulation should not be interpreted as a carte blanche to claim such a system/metric would perform well in real-world settings. On the other hand, academic recommender systems are often developed as generalized rating prediction engines, with no specific platform in mind. Hence, a recommender system failing to perform well in simulation should be taken as strong negative evidence that such a system would fail to perform well in the real world and should probably be reworked; just as a supervised learning algorithm that fails to fit a simple toy problem would be seen with suspicion.

We hope that our simulation framework can be incorporated in the evaluation pipeline of recommender systems since simulations can catch many issues that would otherwise only surface when a recommender system is deployed. When combined with offline evaluation, simulation represents a powerful tool to identify good candidates for deployment. Furthermore, simulation is a useful lens through which to study recommendation algorithms. Our work uncovered many interesting recommender system phenomena that could not have been surfaced through offline evaluation only. It could not have been surfaced in real-world online evaluation either, since it would have been impossible to e.g. switch on and off the user dynamics while keeping everything else fixed.
We believe that simulation studies can uncover many more such phenomena and lead to a more complete understanding of the complex dynamics in recommendations. Especially as we develop simulations that better encapsulate user behaviors, the simulated results provide a window into potential real-world scenarios. For example, Section~\ref{sec:exploration} suggests a significant impact on a recommender's performance from biased sampling, however further work remains to be done to characterize this phenomenon in full generality. 
\begin{figure} \centering
    \includegraphics[width=0.45\textwidth]{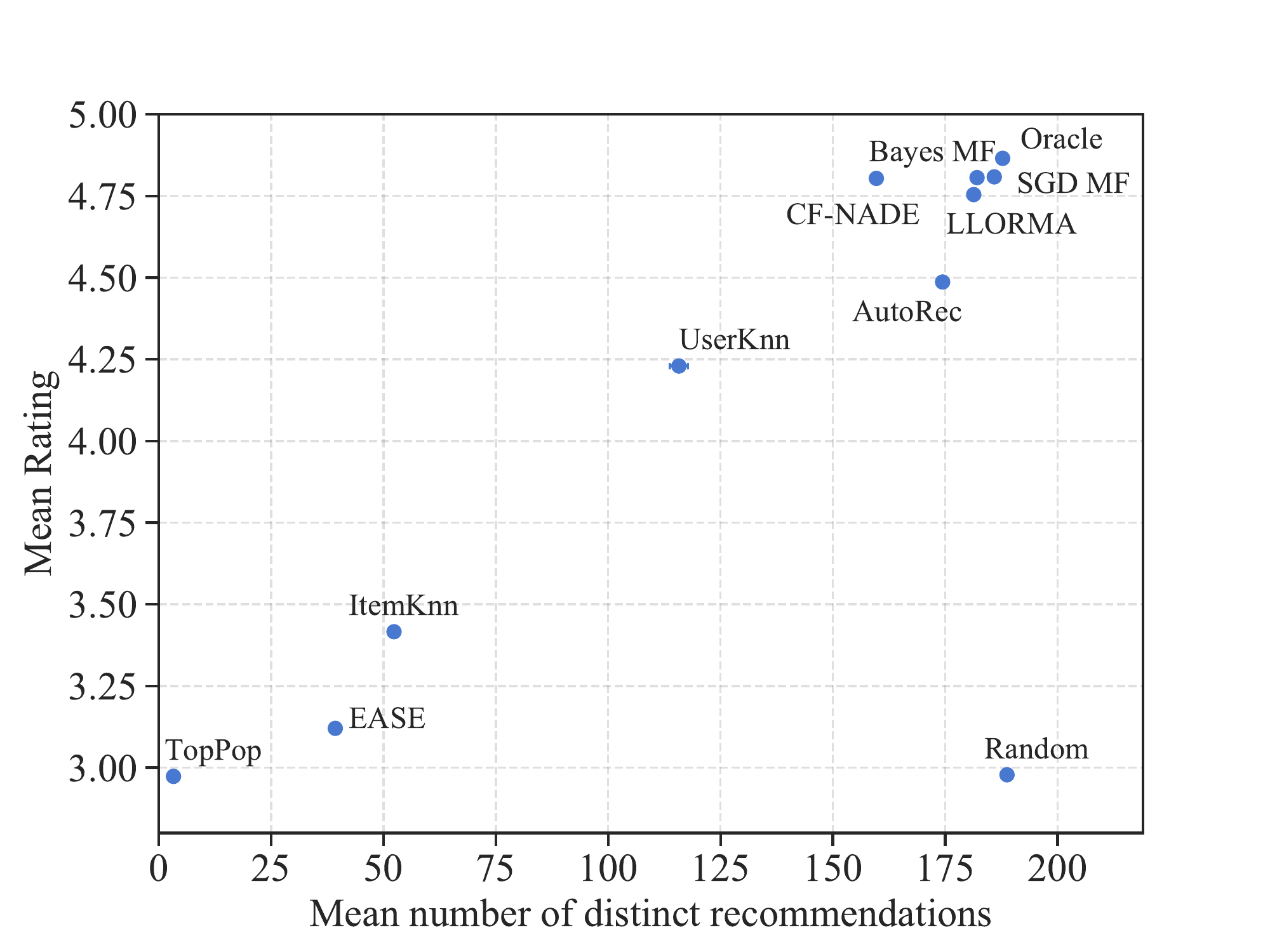}
    \includegraphics[width=0.45\textwidth]{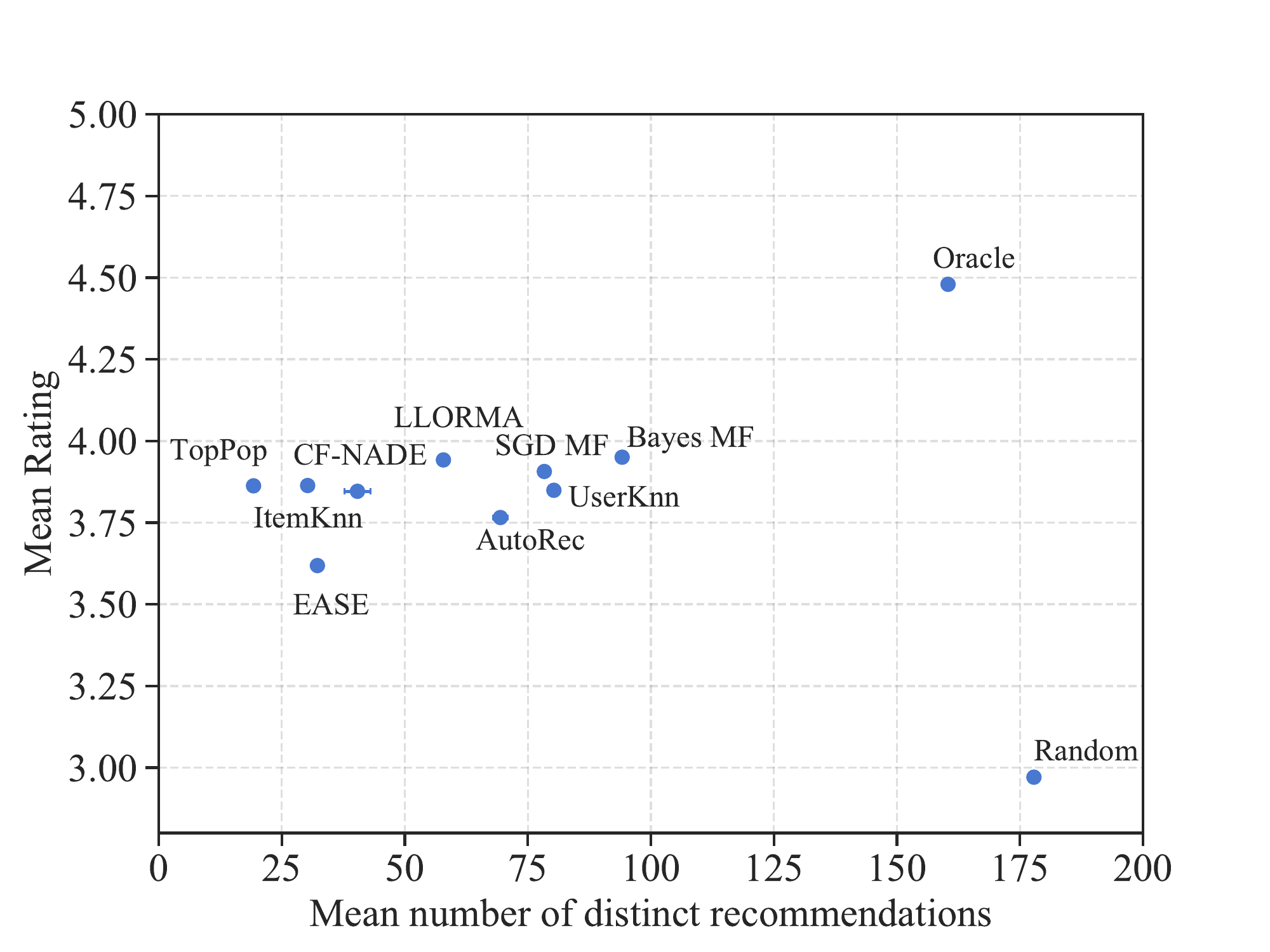}
    \caption{Left: Mean number of distinct items recommended at each timestep plotted against the mean user ratings of all recommended items on the \texttt{topics-static} environment. Right: Mean number of distinct items recommended at each timestep plotted against the mean user ratings of all recommended items on the \texttt{latent-static} environment.}
    \label{fig:coverage}
\end{figure}

\subsection{Diminishing Returns}
Our results provide strong evidence that offline metrics are predictive of online performance. However, this comes with a number of caveats. Increases in offline metrics lead to diminishing returns in terms of online performance, as shown in Figure~\ref{fig:main-plot}. The hesitance of technology companies to adopt expressive yet computationally expensive models \cite{netflixblog} indicates that we may already operate in this regime of diminished returns for real-world recommendation tasks. Furthermore, recommendation tasks customarily involve an abundance of data with which most current algorithms might already be predicting near optimally. Instead, large accuracy gains in these high-data settings could be obtained through the measurement of new user and item features rather than algorithmic innovation.

This issue is compounded by another source of diminishing returns: as predictive models achieve higher-and-higher scores on offline benchmarks, their complexity grows exponentially \cite{openaiblog}. This is not a feasible solution for deployed recommenders where models must handle up to billions of users and items \cite{covington2016deep}. Even our simulations, which are modest in size when compared to a production system, took many thousands of compute hours due to the computational complexity of state-of-the-art models.
These issues indicate that further research effort would be better spent gaining deeper understanding of existing algorithms and datasets to guide the focus of algorithmic improvements. 

\subsection{Metrics Beyond Accuracy}


Though our main motivation was to evaluate algorithmic performance for models trained to maximize accuracy, the data we gathered allows for an analysis of additional metrics.
Here, we briefly describe some observations pertaining to alternate metrics to motivate future in depth analyses,
Figure~\ref{fig:coverage} shows that the relationship between coverage \cite{kaminskaseval} and mean user rating is dependent on the environment.\footnote{We also experimented with the Gini coefficient, a measure of aggregate recomendation diversity, and observed the same results.} The average number of distinct items within a timestep is positively correlated with mean ratings in the \texttt{topics-static} environment, however in \texttt{latent-static} coverage does not correlate with RMSE or mean ratings. We hypothesize that the low correlation in \texttt{latent-static} is due to the presence of item biases, making it easier for recommenders to learn items with high biases rather than exploring long-tail items. Future work may explore the effect of item biases and user behaviors on coverage and diversity.


We hope that these observations will motivate future work to consider how metrics such as fairness, coverage, novelty, and diversity
can be used to design better recommendation systems.
In data-rich settings where the performance gaps between recommenders are minimal, metrics beyond accuracy can guide decision-making about model choice and tuning strategies.
Furthermore, varied metrics may be a useful proxy for accounting for real-world user behavior; while capturing the exact dynamics of evolving user preferences is a challenging task,
ideas of novelty and diversity may help bridge the gap in bettering user experiences in the long run.

\newpage

\bibliography{ref}

\clearpage

\begin{appendices}

\section{The RecLab Framework}
\begin{figure*}[h]
    \centering
    \includegraphics[width=0.75\textwidth]{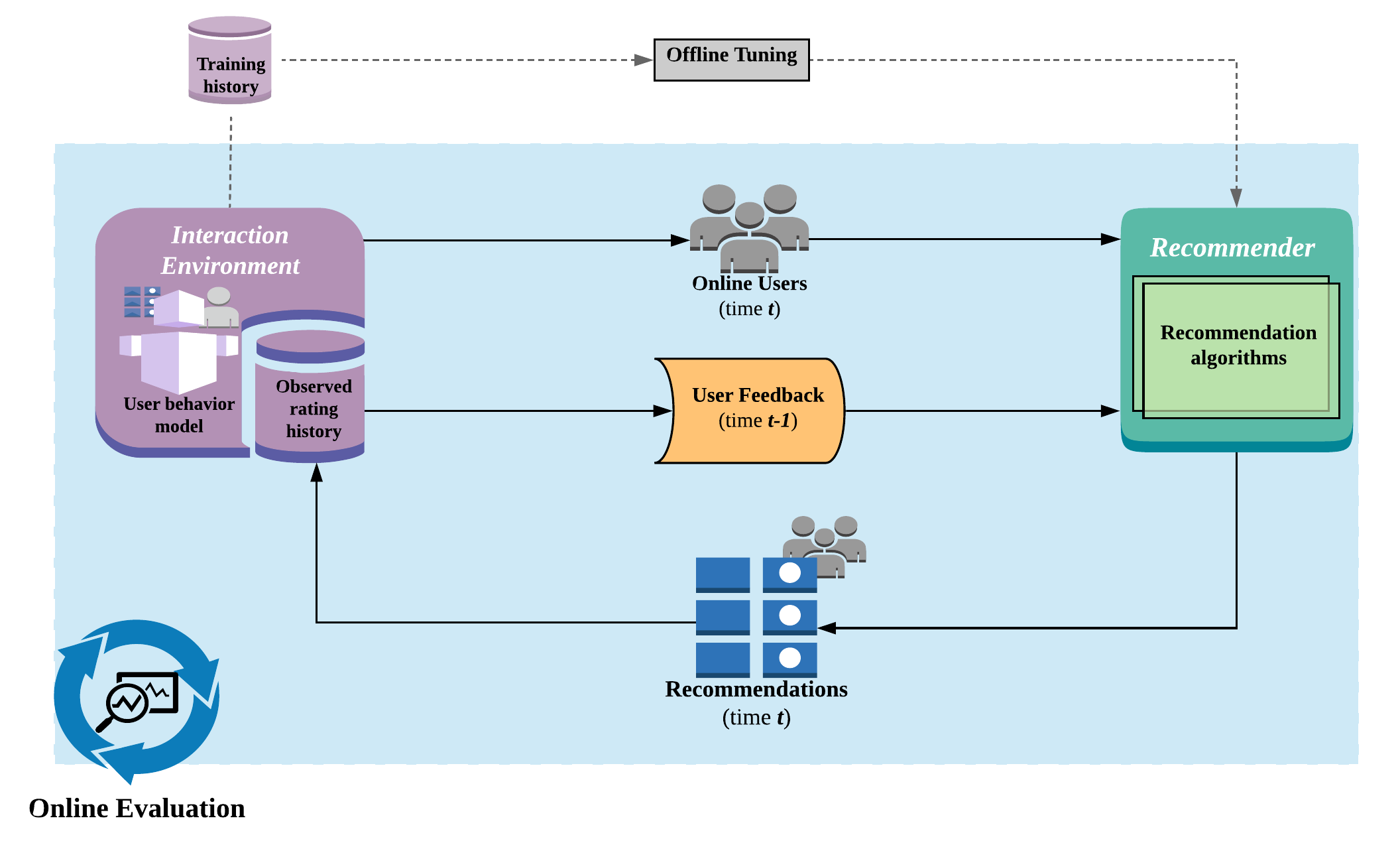}
    \caption{Illustration of the offline-first evaluation and online evaluation pipeline}
    \label{fig:rec_lab}
\end{figure*}

Figure~\ref{fig:rec_lab} shows the process of evaluating a recommendation system within RecLab.

\section{RMSE on \texttt{Topics-static}}
Figure~2 shows the RMSE plotted against the mean rating of all recommended items on the \texttt{latent-static} environment.
\begin{figure}[h]
    \centering
    \includegraphics[width=0.4\textwidth]{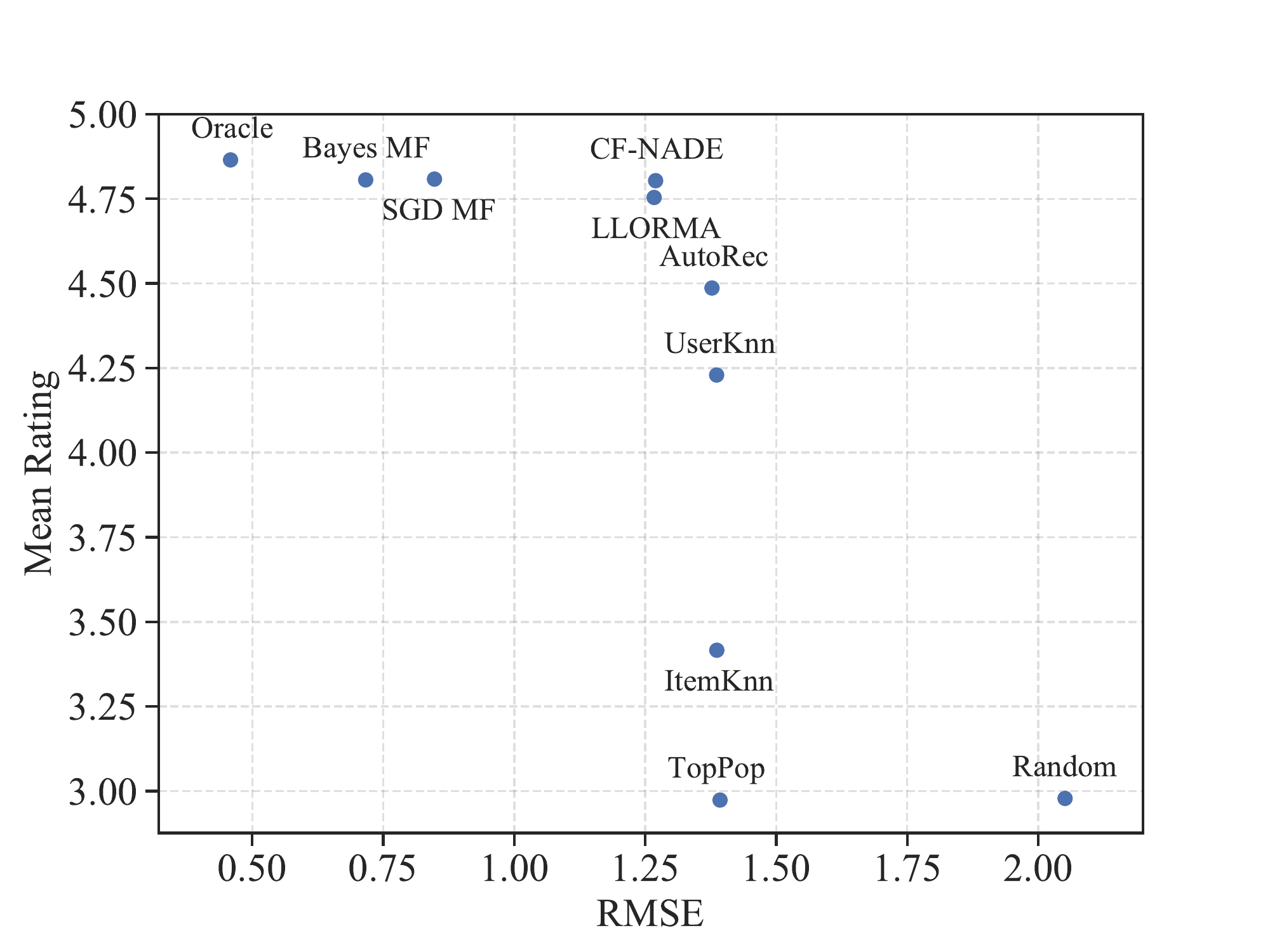}
    \caption{The RMSE plotted against the mean user ratings of all recommended items on the \texttt{topics-static} environment. RMSE is averaged across 5 folds on the offline dataset associated with the environment, user ratings are averaged across 10 trials. Each point represents a single model evaluation with error bars representing 95\% confidence intervals.}
\end{figure}

\section{Exploration on \texttt{Latent-static}}
\begin{figure*}
    \centering
    \includegraphics[width=0.45\textwidth]{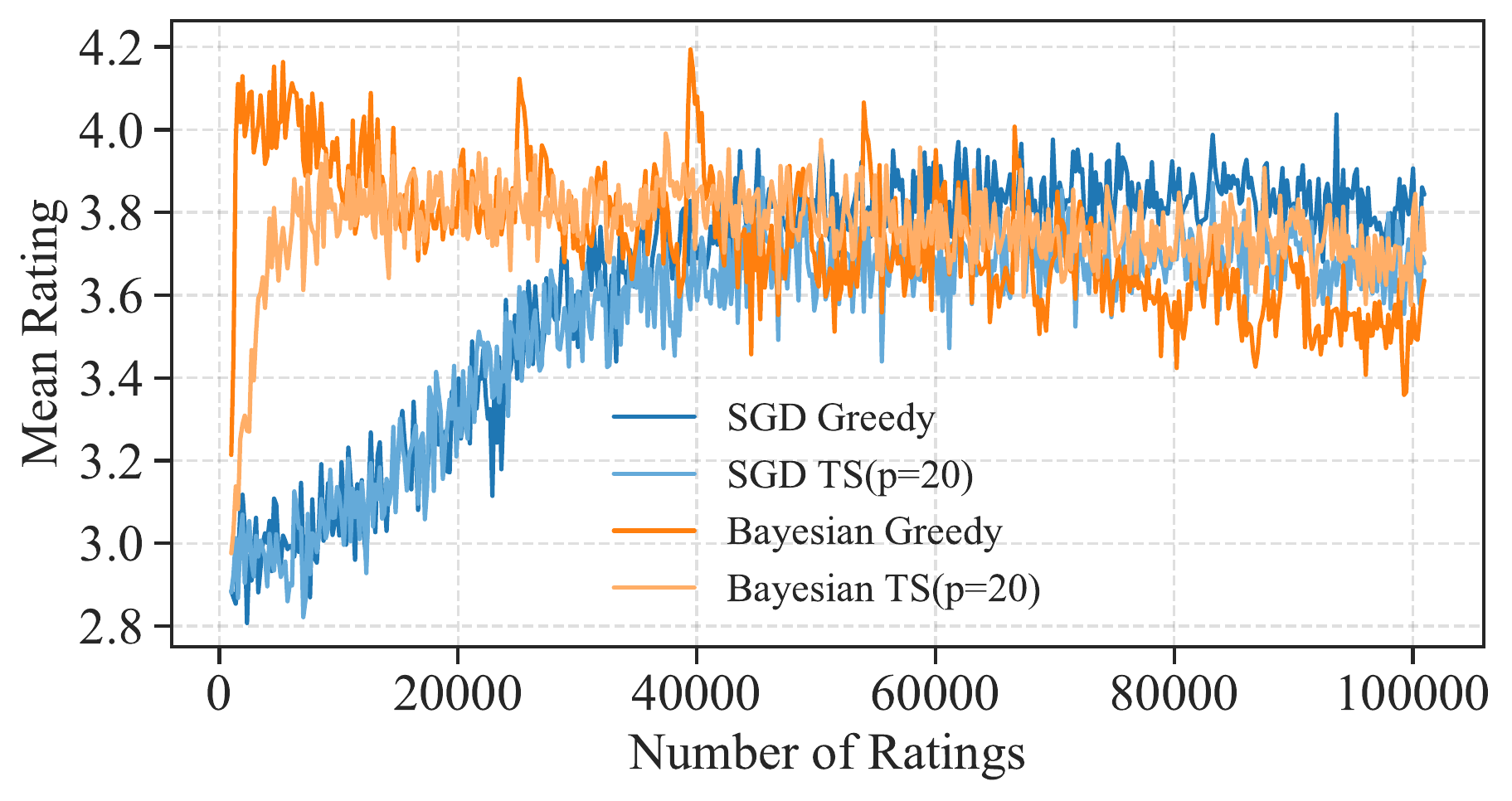}
    \includegraphics[width=0.21
    \textwidth]{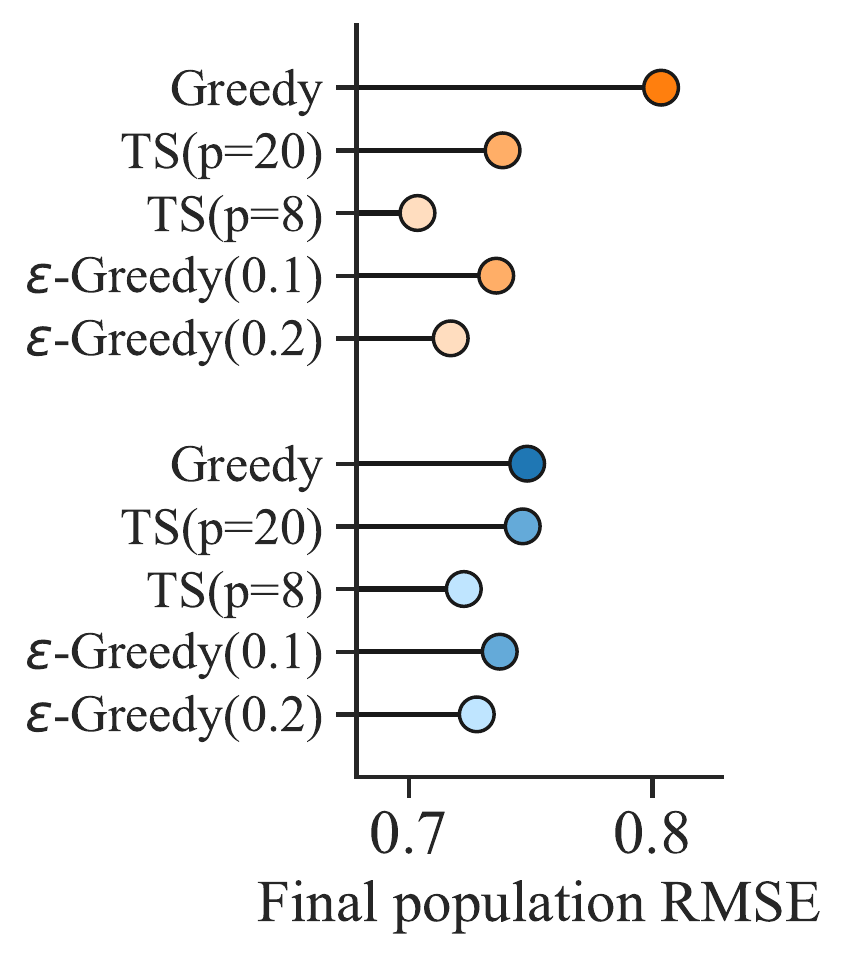}
    \includegraphics[width=0.31\textwidth]{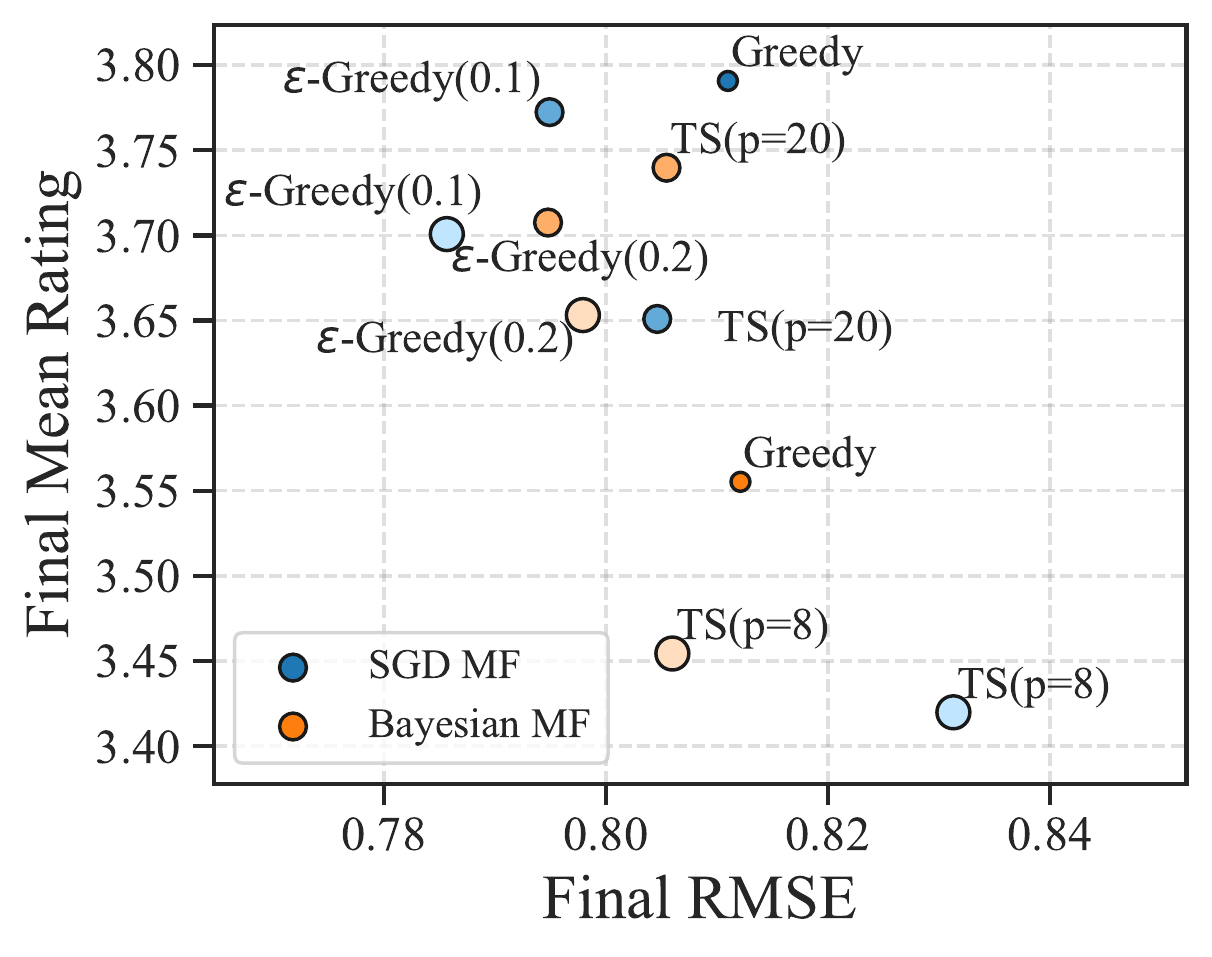}
       \caption{Exploration strategies on \texttt{latent-static-lowdata}. Left: The mean rating over time. Middle: The population RMSE at the end of the experiment measures the overall identification. Right: The final RMSE and mean rating indicate the online performance.
    Final metrics are computed as the average of the 1,000 final ratings.}
    \label{fig:explore_latent}
\end{figure*}

Figure~\ref{fig:explore_latent} shows the effect of exploration on \texttt{latent-static-lowdata}. The observations from the exploration experiments on \texttt{latent-static-topics} are generally corroborated by the results on \texttt{latent-static-lowdata}.
However, there are a few key differences.
First, notice that \texttt{Bayes MF} is able to immediately achieve high mean ratings followed by a decreasing performance.
The \texttt{latent-static} environment generates ratings with an item bias term, in contrast to \texttt{topics-static}, which does not have an inherent quality or popularity structure among the items.
We therefore hypothesize that \texttt{Bayes MF} is exploiting high popularity items at the expense of learning personalized preferences.
We also see a difference in the relative performance of \texttt{SGD MF} strategies: the right panel in Figure~\ref{fig:explore_latent} shows that greedy does not achieve the best final RMSE.

\end{appendices}

\end{document}